\documentclass[11pt,english]{article} 
\usepackage[T1]{fontenc}
\usepackage[latin1]{inputenc} 
\usepackage{graphics}
\usepackage{babel} 
\begin{document} 
\begin{center} 
{\large A modified  Ozer-Taha type cosmological model}

\vspace{2cm}

 { Moncy V.  John$\dag$
 and K.  Babu Joseph},

{\it Department of Physics, Cochin University of Science and Technology,
Kochi 682022,  India}

\noindent {\dag} {\it Department of Physics, St.  Thomas
College,
 Kozhencherri, Kerala, India 689641. }

\vspace{1.5cm}

\vspace{2cm} {\bf Abstract} 
\end{center}

A modified version of the Ozer and Taha nonsingular cosmological model is 
presented on the assumption that the universe's radius is complex if it is 
regarded as empty, but it contains matter when the radius is real. It also 
predicts the values: $\Omega_M =\rho_M/\rho_C\approx 4/3$,  $\Omega_V 
=\rho_V/\rho_C\approx 2/3$, and $\Omega_{-} =\rho_{-}/\rho_C\ll 1$ in the 
present nonrelativistic era, where $\rho_M =$ matter density, $\rho_V$= vacuum 
density, $\rho_{-}$= negative energy density and $\rho_{C}=$ critical density.

\vspace{1.5cm} 
\noindent {PACS 98:80}

\newpage

Inflationary models of the universe \cite{guth123} make use of a scalar field to 
achieve an exponential expansion or inflation at the very early stages of the 
origin of the universe and resolve many of the problems of the standard big bang 
model \cite{kolbturner}. Recently, it was argued \cite{hu,liddle} that inflation 
is the unique mechanism capable of solving the horizon and flatness problems and 
the problem of the generation of density perturbations consistent with  COBE 
observations. Their arguments are only to show that certain features generic to 
all inflationary models, viz., an accelerating scale factor $R(t)$ (which is 
referred to as superluminal expansion) and entropy production are essential 
ingradients in the solution of cosmological problems. Hu, Turner and Weinberg 
\cite{hu} have shown that inflation or something very similar, characterised by 
these two features, may be the only dynamical solution to the horizon and 
flatness problems. Liddle \cite{liddle} has made use of arguments which depend only on the behaviour of scale factor to indicate that the condition $\ddot{R}(t)>0$ during the early phase of the universe is necessary for generating the correct density perturbations, though this condition is valid only in the context of an inflationary explanation of structure formation. It is interesting to note that Hu et. al. or Liddle do not warrant the existence of a scalar field or an exponential expansion, but only an accelerating scale factor along with entropy production. Lidddle precludes even the `coasting evolution' $R(t) \propto t$ as viable.

However, most inflationary models leave certain problems of the big bang model unsolved.  These include the singularity problem, cosmological constant problem etc. The large scale creation of matter at the instant of big bang, violating the conservation of energy - one of the most cherished principles of physics - is referred to as the singularity problem \cite{narpad}. Similarly the smallness of the present vacuum energy density compared to any fundamental scale in physics is not explained in the inflationary models also and is known as the cosmological constant problem \cite{weinpap}. In addition, there are several competing models in inflationary cosmology which differ in the choice of scalar field potential and initial conditions which themselves use physics untested in the laboratory and remain largely speculatory and phenomenological. Hence it is desirable to look for  models of the early universe with an accelerating scale factor and energy production where the above cosmological problems do not appear. One such approach was that of Ozer and Taha \cite{ozertaha} who tried to solve these problems at the level of classical general relativity, without resorting to any scalar field or exponential expansion. Vacuum energy plays a crucial role in this model as a cosmological constant varying as $R^{-2}$. Such time varying cosmological constant models have been investigated by many authors \cite{chen}. From some very general considerations in line with quantum cosmology, Chen and Wu \cite{chen} have argued in favour of an $R^{-2}$-dependence of an effective cosmological constant. By using the Landau-Lifshitz theory for non-equilibrium fluctuations, Pavon \cite{chen} has found that the Ozer-Taha model and the model of Chen and Wu successfully pass their test of thermodynamic correctness. The former model solves the main cosmological problems, but in turn, does not explain the observed near equality of matter density $\rho_M$ of the present universe to the critical density $\rho_C$ (which is referred to as the flatness problem). Instead, this equality is the ansatz they have employed.

In this letter, by postulating the existence of an empty space-time with a complex scale factor, we show that the Ozer and Taha model is obtainable from first principles at a more fundamental level. Their model appears to be a special case of the present one. In this approach, we find that $\Omega_M =\rho_M/\rho_C =1$ in the relativistic era, but in the present non-relativistic era, the same quantity is predicted to have a value very nearly equal to 4/3. The evolution of the scale factor of observed universe is found to be in the same manner as treated in Ref. \cite{ozertaha} to allow for the solution of other cosmological problems and the vacuum energy varies as $R^{-2}$ as prescribed by Chen and Wu. While Abdel-Rahman \cite{chen} has sought to generalize the Ozer-Taha model by using the Chen-Wu prescription under the impression that the critical density assumption of the former is hard to justify, our model derives both.

To start with we note that the Ozer-Taha model can be rederived by assuming that the universe is filled with a total energy density $\tilde{\rho}$ and total pressure $\tilde{p}$, both of which vary with the scale factor 

\begin{equation}
R(t)  = R_{o} e^{\alpha (t)}, \qquad \alpha (0)=0 \label{eq:R}
\end{equation}
as

\begin{equation}
\tilde{\rho}  = \frac{3}{8\pi G}\left[\frac{2}{R^{2}} -\frac{ R^{2}_{0}}{R^{4}}\right]
\end{equation}

\begin{equation}
\tilde{p}  = -\frac{1}{8\pi G}\left[\frac{2}{R^{2}} +\frac{ R^{2}_{0}}{R^{4}}\right]
\end{equation}
with $R=R(t)$. Let us also assume that the space is closed. This means

\begin{equation}
k=+1
\end{equation}
in the Friedman-Robertson-Walker metric. Then the Einstein equations give

\begin{equation}
\frac {\ddot {R}}{R} = -\frac{4\pi G}{3}(\tilde{\rho}+3\tilde{p})= \frac {R^{2}_{0}}{R^{4}},
\end{equation}

\begin{equation}
 \left[ \frac  {\dot {R}}{R} \right] ^{2} + \frac {1}{R ^{2}} = \frac{8\pi G}{3} \tilde{\rho} =\frac {2}{R^{2}} - \frac 
{R^{2}_{0}}{R^{4}}  
\end{equation}
where overdots denote time-derivatives. Solving one gets,

 \begin{equation}
   R^{2} (t) = R^{2}_{0} + t^{2}
   \end{equation}
 which is the result obtained in Ref. \cite{ozertaha}. Furthermore, we assume that $\tilde{\rho}  =  {\rho} _{M} +  {\rho} _{V}$, $ \tilde{p} =  {p}_{M} +  {p}_{V}$ where $\rho_M$ and $\rho_V$ are matter density and vacuum energy density respectively and $p_M$ and $p_V$ are the corresponding pressures. The equations of state are also assumed to have the form ${p}_{M} = w\;  {\rho} _{M}$, ${p}_{V} = -  {\rho} _{V}$. Here, $w=1/3$ for a universe dominated by relativistic matter and $w=0$ when it is dominated by nonrelativistic matter. Unsing these Eqs. (2) and (3) give

\begin{equation}
 {\rho} _{M} = \frac{4}{8\pi G(1+w)}\left[\frac{1}{R^{2}} -\frac{R^{2}_{o}}{R^{4}}\right]
\end{equation}

\medskip

\begin{equation}
 {\rho _{V}}  = \frac{1}{8\pi G(1+w)}\left[ \frac {2(1+3w)}{R^{2}} + \frac {R^{2}_{0}(1-3w)}{R^{4}}\right]
\end{equation}
Eq. (6) may be used to obtain the critical density as 

\begin{equation}
\rho _ {C} \equiv \frac{3}{8\pi G} \dot{\alpha}^2=\frac {3}{8\pi G} \left[ \frac{1}{R^2} -\frac {R_0^2}{R^4}\right].
\end{equation}
Then the ratio of matter density to critical density in the relativistic and nonrelativistic eras can be computed as

\begin{equation}
 {\Omega} _{M,\hbox {rel.}}\equiv \frac { {\rho }_{M,\hbox {rel.}}}{ {\rho }_{C}}=1
\end{equation}

\begin{equation}
 {\Omega} _{M,\hbox {n.rel.}}\equiv \frac { {\rho }_{M,\hbox {n.rel.}}}{ {\rho }_{C}}=\frac {4}{3}
\end{equation}
respectively. Note that (11) is the ansatz in Ref. \cite{ozertaha}. Similarly we get $\rho_{V.rel}=\frac{3}{8\pi G} \frac{1}{R^2}$ as in  Ref. \cite{ozertaha}. But the results obtained for the non-relativistic era are outside the scope of that model.

Now having rederived the Ozer and Taha model, we show that the assumptions (2), (3) and (4) are obtainable from first principles at a more fundamental level. To this end, we start with the FRW metric

\begin{equation}
ds^{2} = dt^{2} - R^{2}(t)\left[\frac{dr^2}{1-kr^2} + r^{2} d\theta ^{2} + r^{2}\sin ^{2}\theta  d\phi ^{2}\right]
\end{equation}
If we make a substitution $R(t) \rightarrow \hat{R}(t) = R(t)e^{i\beta }$ 
in  this metric,,
 the space-time has Lorentzian signature (+ - - -) when $\beta  = \pm n \pi$,  $ ( n = 0,  1,  2, ..)$
and its  signature is Euclidean(++++) when
$\beta  =\pm  (2n +1)\pi /2$, $( n = 0,  1,  2, ..)$. 
Such complex substitutions are not unnatural in relativity. For example, complex substitutions relating open 
and closed isotropic models, de Sitter and anti-de Sitter spacetimes, Kerr and Schwarzchild metrics etc.
are well known  \cite{fla}.
In our case, let $|\hat{R}(t)|=R(t)$  be   in   the     form
$R_{o} e^{\alpha (t)}$ so that
 $\hat{R}(t) = R_{0} e^{\alpha (t)+i\beta }$.
Interesting physics appears if we assume that the time-dependence of
the scale factor is shared by $\beta $ also; i.e., $\beta  = \beta (t)$, 
an assumption consistent with the homogeneity and isotropy scenario \cite{narlikar}.
Our  ansatz   is   to replace $R(t)$ in the FRW metric with

\par

\begin{equation}
\hat{R}(t) = R(t) e^{Z(t)} \equiv  R_{0} e^{\alpha (t)+i\beta (t)}= R(t) e^{i\beta (t)}  
\end{equation}
Further, we  assume that this spacetime isempty with a zero energy-momentum tensor.
Then the   evolution of $\hat{R}(t)$ is dictated by  the  
Einstein-Hilbert action principle, where the action is \cite{kolbturner} 

\par

\begin{equation}
S = \frac{-1}{16\pi G}\int d^{4}x (-g)^{1/2} {\cal R}
\end{equation}
Here ${\cal R}$   is the Ricci scalar  \cite{kolbturner}

\begin{equation}
{\cal R} = \frac{6}{N^{2}}\left[\frac{\dot{\hat{R}}}{\hat{R}}\right]^{2} -\frac{6}{\hat{R}^{2}}
\end{equation}
$N$ is the lapse  function. The Euler-Lagrange equation obtained by minimizing the action with respect to variations of $N$ and $Z$ may be rearranged to get

\begin{equation}
\ddot{Z} +\dot{Z}^2=\frac {\ddot {\hat{R}}}{\hat{R}}=0
\end{equation}

\begin{equation}
\dot{Z}^2+\frac{k}{R_0^2 e^{2Z}}=\left[ \frac 
{\dot {R}}{R} \right] ^{2} + \frac {k}{R ^{2}}=0
\end{equation}
where the gauge $N=1$ is employed. These equations are written for an empty universe with complex scale factor $\hat{R}$ and imply $\ddot {\hat{R}}=0$ and $\dot{\hat{R}}^2+k=0$ respectively. With $\hat{R}(t)=x(t)+i y(t)$, $x$, $y$ real, if we choose $\dot{x}(0)=0$, these give $\dot{y}^2=k$, which implies $k=+1$ (where we rule out the static solution $\dot{x}=0$, $\dot{y}=0$). At $t=0$, if $x=x_0$ and $y=y_0$,

\begin{equation} 
\hat {R}(t) = x_{0} + i\; (y_{0} \pm t)
\end{equation}
Using (14),

\begin{equation}
R(t) = [ x^{2}_{0} + (y_0\pm t)^{2}]^{1/2}
\end{equation}
and

\begin{equation}
\beta (t) = tan ^{-1} ( \frac {y_0\pm t} {x_{0}})
\end{equation}
Note that (21) gives

\begin{equation}
\dot {\beta }(t) = \frac {\pm x_{0}}{R^{2}(t)} = \frac {\pm cos ^{2} \beta }{x_{0}}
\end{equation}
If we choose the origin of time such that $y=0$ at $t=0$, then $y_0=0$. Relabelling $x_0=R_0$, (20) gives the solution (7). Now using (20)-(22) and $\alpha =\log \left[\frac{R(t)}{R_0}\right]$ in the real parts (17) and (18) give (5) and (6). Assumptions (2) and (3) naturally follow. Thus the complex quantity $\hat{R}$ is the scale factor of an empty expanding space, which we call a hidden universe whereas $R(t)=\mid \hat{R}(t)\mid $ is the scale factor of a closed, expanding space with total energy density $\tilde{\rho}$. As shown in Ref. \cite{ozertaha}, the latter is an acceptable model of the observed universe. If we had chosen $\dot{y}=0$, $\dot{x}^2=-1$ in the solution of (17) and (18), eventhough we get $k=-1$ for the hidden universe, solution (7) and Eqs. (5) and (6) may be regained by a similar procedure, so that the observed universe with total energy density $\tilde{\rho}$ is again closed, which is assumption (4).

However, we note that the predictions (8) and (9) are not the most natural ones. Eq. (8) gives $\rho_M=0$ at $t=0$. Simialrly from Eq. (9), we see that $\rho_V$ has an $R^{-4}$ dependence in the nonrelativistic era. Chen and Wu \cite{chen} have argued in favour of an $R^{-2}$ behaviour for $\rho_V$ on the basis of some dimensional considerations in line with quantum cosmology. In order to avoid these less probable results mentioned above, we suggest an alternative ansatz,

$$
\tilde{\rho}  =  {\rho} _{M} +  {\rho} _{V} +  {\rho} \_
 , \qquad
\tilde{p} =  {p}_{M} + { p}_{V} +  {p}\_
$$

\begin{equation}
{p}_{M} = w\;  {\rho} _{M},
\qquad
 {p}_{V} = -  {\rho} _{V},
\qquad
 {p}\_ = \frac{1}{3}  {\rho} \_
\end{equation}
and

$$
 {\rho} \_ = - \frac{3}{8\pi G}\frac{R^{2}_{o}}{R^{4}},
$$
where $\rho_-$ and $p_-$ are the energy density and pressure respectively, which are appropriate for negative energy relativistic particles. Negative energy densities in the universe were postulated earlier \cite{narpad}. Solving (2) and (3) with (23), we get, instead of (8) and (9),

\begin{equation}
 {\rho}_{M}  =\frac{4}{8\pi G(1+w)}\frac{1}{R^{2}}
\end{equation}

\begin{equation}
 {\rho}_{V}  =\frac{2(1+3w)}{8\pi G(1+w)}\frac{1}{R^{2}}
\end{equation}
This avoids the afore-mentioned impasse. Thus our predictions for $R(t)\gg R_0$ are
$ {\Omega} _{M,\hbox{rel.}} \approx  1, \qquad  {\Omega} _{V,\hbox{rel.}} \approx  1, \qquad
 {\Omega} _{M,\hbox{n.rel.}} \approx 4/3, \qquad  {\Omega} _{V,\hbox{n.rel.}} \approx  2/3
$ and ${\Omega}\_\equiv { {\rho}\_}/{\rho_{C}} \ll 1 $. It is easy to see that the conservation law for total energy is obeyed, irrespective of either of the ansatze regarding the detailed structure of $\tilde{\rho}$ and $\tilde{p}$.

The solution of cosmological problems as given in \cite{ozertaha} can be performed in this model also if we associate thermal radiation at  temperature $T$  with
energy densities ${\rho} _{M}$  and $ {\rho} \_$ as 

\par

\begin{equation}
 {\rho}_{M}  +{\rho} \_ = \frac{1}{30} \pi ^{2} N(T) T^{4}
\end{equation}

\noindent where $N(T)$ is the effective number
 of spin degrees  of  freedom   at
temperature $T$ and the units are such that $\hbar = c = k_{B} = 1$. In addition, the present model explains the near equality of matter density and critical density in a natural way.

Since   the   expansion
process is reversible and the basic equations   are  time-reversal
invariant,  we can extrapolate to $t <0$ to obtain   an   earlier
contracting phase  for the universe. Such an earlier phase  was  proposed  by
Lifshitz and Khalatnikov \cite{lif}. If there was such  an  initial  phase,
causality  could  have  established itself much earlier   than   the
time   predicted in \cite{ozertaha}.

The  new model predicts creation  of  matter  at   present
with a rate of creation per unit volume given by

\par

\begin{equation}
\frac{1}{R^{3}}\frac{d(R^{3} {\rho} _{M})}{dt} \mid_{P}\; =  {\rho} _{M,P}\;\dot{\alpha} _{P}
\end{equation}
where $ {\rho} _{M,P}$  is the present matter density and $\dot{\alpha}_P$ is the present value of Hubble's constant. 
In arriving at this  result,   we   have
made  use of the assumption that the present universe is dominated by nonrelativistic  matter. The required rate of  creation  is
only  one-third of that in the steady state cosmology \cite{narlikar}. Such a
possibility of creation of matter 
cannot be ruled out at the present level of observation.

To summarize, we note that we have a model for the universe free from all cosmological problems. We started with the assumption that the universe's radius is $\hat {R}$, a complex quantity and that it contains no energy density. But when we regard the radius to be $\mid \hat{R}\mid$, a real quantity, then the universe appears to be non-empty. We call the former a hidden universe whereas the latter is seen to be an acceptable model of the observed universe.  The energy density of the observed universe is composed of matter density, vacuum energy density, and negative energy density. The minimum radius  $R_0$ is not fixed by the theory. The various predictions of the model are made under the assumption that $R_0 \ll t_P$, the present age of the universe. These predictions are well within the range of observed values, except for the value of $\Omega_M$. A possible existence of dark matter in the universe may explain this discrepancy. In particular, using (7), the model correctly predicts the value of the combination
 $(\frac{\dot{R}}{R})_P \;t_P = \dot{\alpha}_P \;t_P \approx 1$, verified in all recent observations \cite{jcc}. To be more precise, this value lies in the range

$$
0.85 < \dot{\alpha}_P \;t_P <1.91
$$
as quoted by Lima and Trodden \cite{chen} on the basis of the above observations and places the present approach in a more advantageous position than the standard flat FRW and inflationary models where this value is only 2/3. (This is referred to as the age problem in these models.) The new quantity $\beta (t)$ is viewed as a phase factor. The model is still in the realm of classical general relativity and solves all the cosmological problems without a scalar field.

We thank the referee for valuable comments and for suggesting useful references.

\end{document}